\documentclass[12pt]{iopart}
\usepackage{iopams}

\usepackage{graphicx}

\usepackage{color}

\begin{document}


\title[M E Macovei]{Effect of pressure and Ir substitution in YbRh$_{2}$Si$_{2}$}

\author{M E Macovei, M Nicklas, C Krellner, C Geibel and F Steglich}

\address{Max Planck Institute for Chemical Physics of Solids, N{\"o}thnitzer Str. 40, 01187 Dresden, Germany}

\eads{macovei@cpfs.mpg.de}

\begin{abstract}

In this article we present a study of the electrical resistivity of Yb(Rh$_{1-x}$Ir$_{x}$)$_{2}$Si$_{2}$, $x=0.06$, under high pressure and in magnetic field. Ir substitution is expanding the unit cell and leads to a suppression of the antiferromagnetic transition temperature to zero, where eventually a quantum critical point (QCP) exists. We applied hydrostatic pressure to reverse the effect of substitution. Our results indicate that Yb(Rh$_{0.94}$Ir$_{0.06}$)$_2$Si$_2$ is situated in the immediate proximity to a volume controlled QCP, but still on the magnetically ordered side of the phase diagram. The temperature - pressure phase diagram of Yb(Rh$_{0.94}$Ir$_{0.06}$)$_2$Si$_2$ resembles that of the pure compound. Substitution acts mainly as chemical pressure. Disorder introduced by substitution has only minor effects.

\end{abstract}



\date{\today}

\section{Introduction}

The unique properties that develop around a quantum critical point (QCP) are a major topic of current solid state research. YbRh$_2$Si$_2$ exhibits a weak antiferromagnetic (AFM) transition at atmospheric pressure with a N\'{e}el temperature of only $T_N\approx70$~mK \cite{1}. By applying a small magnetic field perpendicular to the crystallographic $c$ axis the transition temperature can be continuously suppressed to zero at about $B_{c} = 60$~mT ($B_{c}=660$~mT for $B\parallel c$), driving the system to a QCP \cite{2}. The temperature dependence of the specific heat and resistivity reveal an extended non-Fermi-liquid (NFL) regime around the magnetic field-induced QCP \cite{3}. Under external pressure the N\'{e}el temperature is continuously increasing, typical for an Yb-based intermetallic system. The pressure effect on YbRh$_2$Si$_2$ has been intensively studied \cite{4, 5, 6, 7}. YbRh$_2$Si$_2$ can be tuned to the paramagnetic side of the pressure (volume) controlled QCP by increasing the unit-cell volume. This can be achieved only by chemical substitution. Replacing a nominal concentration of $5$ at.\% Si by Ge in YbRh$_2$Si$_2$ leads to a shift of $T_{N}$ from $70$~mK in the pure compound to $T_{N} \approx 20$~mK \cite{3}. A similar result has been reported on small La substitution \cite{8}. Expanding the crystal lattice of YbRh$_2$Si$_2$ by substituting Rh with the isovalent Ir, allows one to tune the system through the QCP without significantly affecting the electronic properties. Recent measurements of the magnetic susceptibility on Yb(Rh$_{1-x}$Ir$_{x}$)$_{2}$Si$_{2}$ demonstrate that for low Ir doping, $x\lesssim 0.025$, the system orders magnetically, while in the crystals with $17$ at.\% Ir substitution, no magnetic transition can be observed \cite{9}. In this work we studied Yb(Rh$_{0.94}$Ir$_{0.06}$)$_{2}$Si$_{2}$ supposed to be at the border of magnetism, by means of electrical resistivity measurements as a function of both hydrostatic pressure ($p$) and magnetic field ($B$). The substitution of $6$ at.\% Rh by Ir leads to a lattice expansion of only about 0.03\%. Our results provide evidence that Yb(Rh$_{0.94}$Ir$_{0.06}$)$_{2}$Si$_{2}$ is situated in proximity to the QCP, but still slightly on the magnetic side of the temperature-volume phase diagram. The $T-p$ phase diagram resembles that of the stoichiometric YbRh$_2$Si$_2$, considering a shift by a rigid pressure corresponding to the lattice expansion due to the $6$ at.\% Ir substitution.

\begin{figure}[b]
\centering
\includegraphics[angle=0,width=90mm,clip]{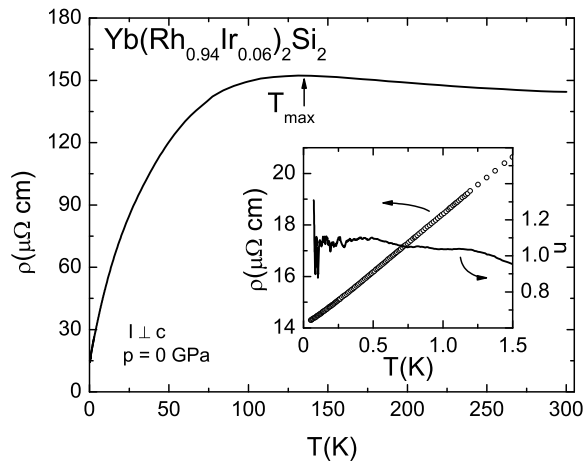}
\caption{Electrical resistivity of Yb(Rh$_{0.94}$Ir$_{0.06}$)$_{2}$Si$_{2}$ at atmospheric pressure measured perpendicular to the $c$ axis as function of temperature. Inset: resistivity (left axis) and temperature exponent $n={\rm d}\ln(\rho-\rho_{0})/{\rm d}(\ln T)$ in the temperature range ${\rm 50~mK} \leq T \leq {\rm 1.5~K}$.\label{figure1a}}
\end{figure}

\section{Experimental details}

Single crystals of Yb(Rh$_{0.94}$Ir$_{0.06}$)$_{2}$Si$_{2}$ were grown from In-flux. The sample stoichiometry has been verified by energy dispersive X-ray analysis (EDX). The tetragonal ThCr$_2$Si$_2$ crystal structure has been confirmed by X-ray powder diffraction. Measurements of the electrical resistivity have been performed using a standard four-point technique at temperatures $50$~mK $\leq T \leq 300$~K and in magnetic fields up to $B=8$~T in a physical property measurement system (Quantum Design) and in a $^3$He/$^4$He dilution refrigerator. The current was applied within the $a-b$ plane and the magnetic field parallel to the crystallographic $c$ axis. In the low-pressure region a piston-cylinder type pressure cell capable of pressures up to $p=3$~GPa with silicone fluid as pressure transmitting medium was used. For pressures $p\leq10$~GPa, a Bridgman-type pressure cell with steatite as pressure transmitting medium was utilized. The pressure inside the pressure cell was determined by monitoring the pressure dependence of the superconducting transition temperature of Sn or Pb, respectively, placed near the sample inside the pressure cell.

\begin{figure}[b]
\centering
\includegraphics[angle=0,width=90mm,clip]{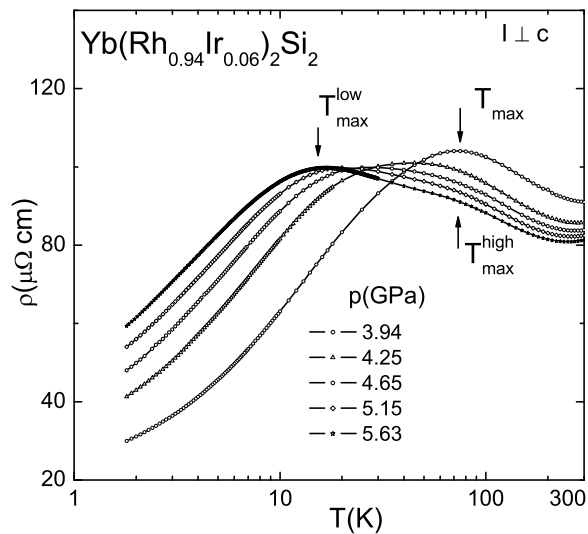}
\caption{Temperature dependence of the electrical resistivity for Yb(Rh$_{0.94}$Ir$_{0.06}$)$_{2}$Si$_{2}$ at different pressures in the temperature range ${\rm 1.8~K} \leq T \leq {\rm 300~K}$. Arrows indicate $T_{\rm max}$ for $p=3.94$~GPa. For $p=5.63$~GPa a maximum at $T^{\rm low}_{\rm max}$ and a shoulder at $T^{\rm high}_{\rm max}$ are clearly distinguishable, both indicated by arrows.\label{figure1}}
\end{figure}

\begin{figure}[t]
\centering
\includegraphics[angle=0,width=65mm,clip]{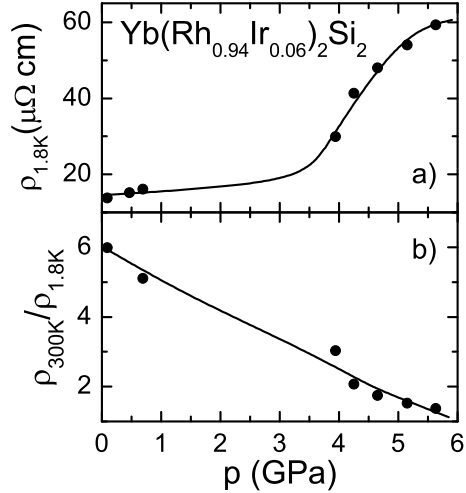}
\caption{Upper panel: isothermal resistivity at $T=1.8$~K, $\rho_{\rm 1.8\,K}$, as function of pressure; lower panel: pressure dependence of the resistivity ratio $\rho_{\rm 300\,K}/\rho_{\rm 1.8\,K}$. The lines are guides to the eye.\label{figure2a_2b}}
\end{figure}

\section{Results and discussion}

Figure~\ref{figure1a} shows $\rho(T)$ at ambient pressure in the temperature range ${\rm 50~mK} \leq T \leq {\rm 300~K}$. The electrical resistivity temperature dependence, $\rho(T)$, of Yb(Rh$_{0.94}$Ir$_{0.06}$)$_{2}$Si$_{2}$ at atmospheric pressure follows the typical behavior expected for a Kondo-lattice system.  Below $300$~K the resistivity increases slightly with decreasing temperature, then exhibits a broad maximum around $T_{max}\approx 135$~K. Upon further cooling $\rho(T)$ strongly decreases due to the onset of coherent Kondo scattering. In the stoichiometric compound YbRh$_2$Si$_2$, the resistivity maximum at $p=0$ is reported at about the same temperature \cite{1}. At low pressure $p\lesssim 4$~GPa, the resistivity shows a single broad maximum around $T_{max} = 100$~K (figure~\ref{figure1}). Upon increasing pressure the maximum shifts to lower temperatures, indicating a decrease of the hybridization between the Yb $4f$ and the conduction electrons. For pressures larger than $p\approx4$~GPa, a shoulder is developing next to the maximum. At $p=4.5$~GPa the maximum is observed at $T^{\rm low}_{\rm max}\approx 45$~K and the shoulder at $T^{\rm high}_{\rm max} \approx 95$~K. The pressure responses of the maximum and the shoulder are different: $T^{\rm low}_{\rm max}(p)$ shifts to lower temperatures upon increasing pressure, while $T^{\rm high}_{\rm max}(p)$ is nearly pressure independent. In pure YbRh$_2$Si$_2$ a similar behavior is found, the single maximum at low pressure splits into two at about the same pressure \cite{6}. The single maximum in $\rho(T)$ at low pressures can be explained by a combination of scattering processes on the ground state doublet and on the excited crystalline electric field (CEF) levels. Taking into account the CEF level scheme obtained from inelastic neutron scattering for YbRh$_2$Si$_2$ at ambient pressure \cite{10}, for $p\geq4.5$~GPa the high temperature shoulder can be attributed to inelastic Kondo scattering on the excited CEF levels and the low temperature maximum to Kondo scattering on the ground state doublet. The very similar pressure evolution of the high-temperature resistivity in YbRh$_2$Si$_2$ and Yb(Rh$_{0.94}$Ir$_{0.06}$)$_{2}$Si$_{2}$ indicates that the pressure effect on the CEF levels is comparable in both materials. However, in Yb(Rh$_{0.94}$Ir$_{0.06}$)$_{2}$Si$_{2}$ the resistivity maximum is situated at slightly higher temperatures compared with YbRh$_2$Si$_2$. The isothermal pressure dependence of the resistivity at $T=1.8$~K, $\rho_{\rm 1.8\,K}(p)$, stays initially nearly constant with increasing pressure before it strongly increases, above $p\approx4$~GPa, by a factor of more than $3$ (figure~\ref{figure2a_2b}a). At the same time the resistivity ratio $RR_{\rm1.8\,K}=\rho_{\rm 300\,K}/\rho_{\rm 1.8\,K}$ decreases monotonically. This reveals that the increase of $\rho_{\rm 1.8\,K}(p)$ above $p\approx4$~GPa is caused by additional incoherent scattering at low temperatures. A pressure-induced increase of the residual resistivity, $\rho_0$, was previously found in different  Yb- and Ce-based compounds, like YbIr$_2$Si$_2$ \cite{Yuan06}, YbCu$_2$Si$_2$ \cite{Alami98}, or CeCu$_2$(Si$_{1-x}$Ge$_x$)$_2$ \cite{Yuan03}. Different mechanisms based on magnetic or valence transitions can lead to a strongly elevated $\rho_0$. It was found theoretically that in quantum-critical systems impurity scattering can be strongly enhanced by quantum-critical spin-fluctuations \cite{Rosch99,Miyake02a}. In a theoretical model based on valence-fluctuations a strong increase of $\rho_0$ and a linear temperature dependence of the resistivity at low temperatures is predicted above a crossover temperature $T_v$ at a valence transition \cite{Miyake02b}. From the present data it is not possible to decide which mechanism leads to the enhanced $\rho_0$ in Yb(Rh$_{0.94}$Ir$_{0.06}$)$_2$Si$_2$.

\begin{figure}[t]
\centering
\includegraphics[angle=0,width=90mm,clip]{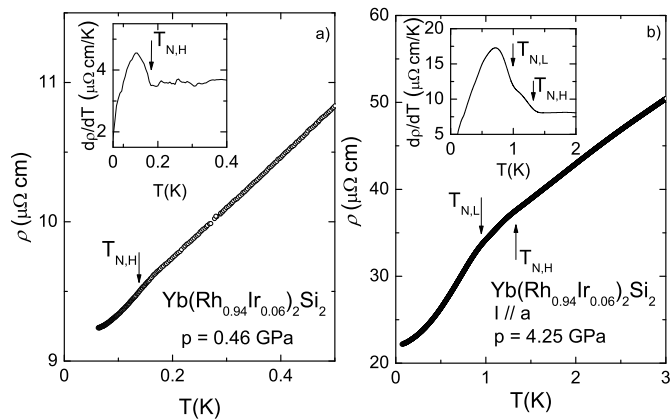}
\caption{Low-temperature electrical resistivity of Yb(Rh$_{0.94}$Ir$_{0.06}$)$_{2}$Si$_{2}$ at a) $p=0.46$~GPa and b) $p=4.25$~GPa. The transition temperatures were determined from the temperature derivative of ${\rm d}\rho(T)/{\rm d}T$ as shown in the insets of panel a) and b) for $p=0.46$~GPa and $4.25$~GPa, respectively. The arrows indicate the magnetic transitions at $T_{N,H}$ and $T_{N,L}$, respectively.\label{figure3a_3b}}
\end{figure}

At atmospheric pressure, the low-temperature resistivity of Yb(Rh$_{0.94}$Ir$_{0.06}$)$_{2}$Si$_{2}$ shows no anomaly pointing to the existence of a magnetic transition in the temperature range down to $T=50$~mK (inset figure 1). In magnetic susceptibility measurements a clear magnetic transition was observed for Yb(Rh$_{0.975}$Ir$_{0.025}$)$_{2}$Si$_{2}$ at $T_{N}\thickapprox40$~mK, while for a sample with $6$ at.\% Ir doping no magnetic transition anomaly could be resolved at temperatures down to $0.02$~K, suggesting that Yb(Rh$_{0.94}$Ir$_{0.06}$)$_{2}$Si$_{2}$ is very close to the QCP \cite{9}. Finally, Yb(Rh$_{0.83}$Ir$_{0.17}$)$_{2}$Si$_{2}$ is on the paramagnetic side of the QCP \cite{9}. Applying pressure of only $p = 0.46$~GPa on Yb(Rh$_{0.94}$Ir$_{0.06}$)$_{2}$Si$_{2}$ is sufficient to shift the AFM transition up to $T_{N,H}\approx 0.14$~K. $T_{N,H}$ is clearly resolved as a kink in $\rho(T)$ (cf. figure~\ref{figure3a_3b}a). A similar feature at the AFM transition has been observed in YbRh$_2$Si$_2$ \cite{1}. With increasing pressure, $T_{N,H}(p)$ shifts to higher temperatures as expected for an Yb-based heavy fermion compound, but the signature of the transition in $\rho(T)$ is becoming less pronounced (figure~\ref{figure3a_3b}b). At $p = 4.25$~GPa, a second more pronounced anomaly is appearing below $T_{N,H}$ at $T_{N,L}=1$~K (indicated by an arrow in figure~\ref{figure3a_3b}b). Two successive magnetic transitions have been also reported in YbRh$_{2}$Si$_{2}$ under pressure \cite{7}.

\begin{figure}[t]
\centering
\includegraphics[angle=0,width=80mm,clip]{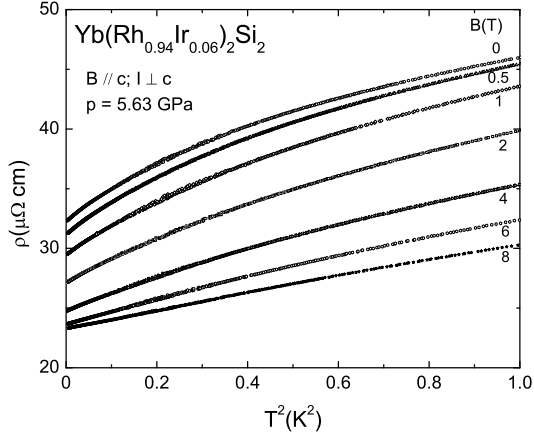}
\caption{Temperature dependence of the resistivity as $\rho$ vs. $T^{2}$ at $p = 5.63$~GPa for different magnetic fields applied parallel to the direction of the crystallographic $c$ axis.\label{figure4}}
\end{figure}

Electrical resistivity at low temperature follows a power-law dependence which can be expressed by $\rho(T)=\rho_0+A_n T^n$. As displayed in the inset of figure~\ref{figure1a}, the temperature exponent $n={\rm d}\ln(\rho-\rho_{0})/{\rm d}(\ln T)$, remains nearly constant below $T=1.5$~K with a value of $n\approx 1\pm 0.1$. A recovery of a $\Delta\rho(T)\propto T^{2}$ behavior is not observed down to the lowest accessible temperature in our experiment. The resistivity exhibits a quasi-linear temperature dependence for all investigated pressures above $T_{N,H}$ characteristic of NFL behavior similar to that observed in the YbRh$_2$Si$_2$. This suggests that disorder introduced by the substitution of Ir for Rh does not affect the quantum critical behavior strongly for this low Ir concentration. However, below the transition temperature $T_{N,L}$, $\rho(T)$ can not be described by a $T^{2}$ dependence as in YbRh$_2$Si$_2$ \cite{3}.

\begin{figure}[t]
\centering
\includegraphics[angle=0,width=75mm,clip]{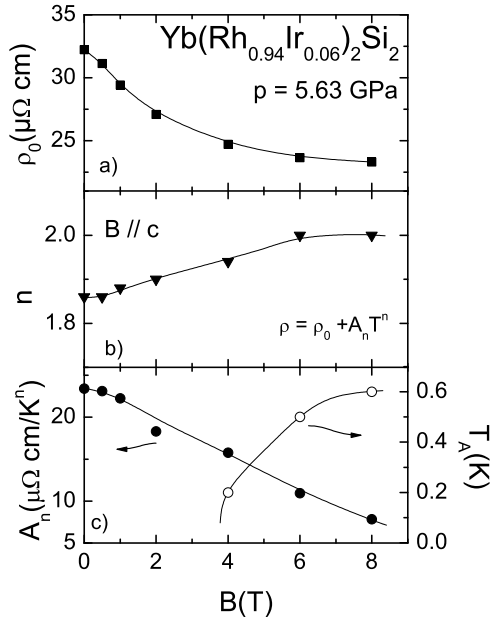}
\caption{Magnetic field dependence of the residual resistivity $\rho_0$ (a), temperature exponent $n$ (b), pre-factor $A_{n}$ (c, left axis) obtained from a fit of $\rho=\rho_0 +A_nT^n$ to the low temperature resistivity data and $T_{A}$ (c, right axis) upper limit of $T^{2}$ dependence of the resistivity at $p = 5.63$~GPa. \label{figure5a_5b_5c}}
\end{figure}

At selected pressure, electrical resistivity has been measured in applied magnetic field. Figure~\ref{figure4} shows $\rho(T)$ of Yb(Rh$_{0.94}$Ir$_{0.06}$)$_{2}$Si$_{2}$ as function of $T^2$ in different magnetic fields for $p = 5.63$~GPa. At this pressure in zero magnetic field $T_{N,H}=1.61$~K and $T_{N,L} = 0.9$~K. Above $T_{N,H}$, $\rho(T)$ follows a quasi-linear temperature dependence. It is interesting to note that at $B=2$\,T, an anomaly can be still observed at about $1.3$~K, but at $B = 8$\,T no indication of any feature is visible in ${\rho(T)}$ anymore implying that magnetic order is suppressed at this magnetic field. In magnetic field the transition anomaly is broadened and, therefore, a complete analysis of the magnetic field dependence of the transition temperature is difficult. The resistivity data below $T\lesssim 0.6$~K can be described by a power-law behavior, $\rho(T)=\rho_0+A_n T^n$. The magnetic field dependence of $\rho_0$, $n$ and temperature coefficient $A_{n}$ for $p = 5.63$ GPa is plotted in figure~\ref{figure5a_5b_5c}. The residual resistivity, $\rho_0(B)$, is monotonically decreasing upon increasing magnetic field, but tends to saturate at large fields ($B\approx 6-8$~T). While in small magnetic fields, $B\lesssim2$~T, a temperature exponent $n\approx1.8$ significantly smaller than $n = 2$ as expected for a Landau Fermi liquid (LFL), is found, in magnetic fields $B\gtrsim4$~T the characteristic behavior of a LFL is recovered. The LFL region is growing in temperature with increasing magnetic field as can be seen by the extended linear region in the $\rho(T)$ {\it vs.} $T^2$ plot for large magnetic fields in figure~\ref{figure4}. The field dependence of the crossover temperature giving the upper limit of the temperature range where the data can be described by a $T^{2}$ dependence is displaced in the figure~\ref{figure5a_5b_5c}c.

\begin{figure}[h]
\centering
\includegraphics[angle=0,width=90mm,clip]{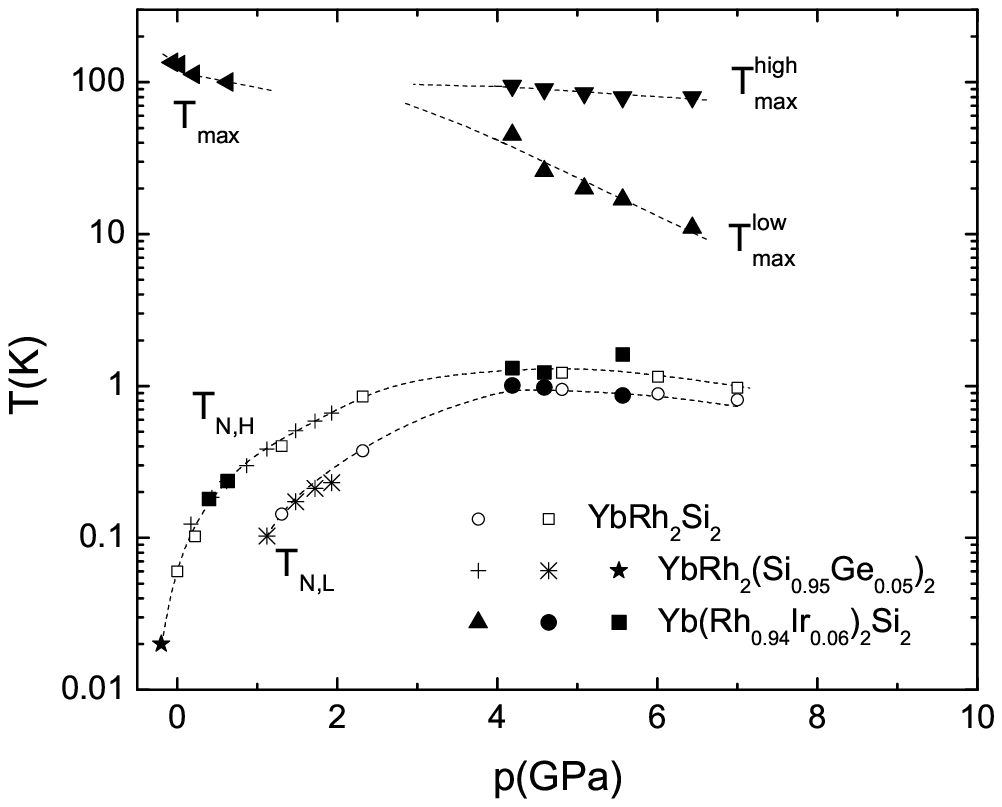}
\caption{Temperature - pressure ($T-p$) phase diagram of Yb(Rh$_{0.94}$Ir$_{0.06}$)$_{2}$Si$_{2}$ (solid symbols). Data of YbRh$_2$Si$_2$ (open symbols, Ref.\,\cite{1,7}) and YbRh$_2$(Si$_{0.95}$Ge$_{0.05}$)$_2$ (crosses, Ref.\,\cite{4}) have been included. $(\star)$:~$T_{N,H}$ of YbRh$_2$(Si$_{0.95}$Ge$_{0.05}$)$_2$ at atmospheric pressure obtained from specific heat measurements (Ref.\,\cite{3}). The data for Yb(Rh$_{0.94}$Ir$_{0.06}$)$_{2}$Si$_{2}$ and YbRh$_2$(Si$_{0.95}$Ge$_{0.05}$)$_2$ have been shifted by a fixed pressure of $p=-0.06$~GPa and $p = - 0.2$~GPa, respectively, with respect to YbRh$_2$Si$_2$.\label{figure6}}
\end{figure}

The $T-p$ phase diagram in figure~\ref{figure6} summarises the results obtained for Yb(Rh$_{0.94}$Ir$_{0.06}$)$_{2}$Si$_{2}$. In addition, data for YbRh$_{2}$Si$_{2}$ \cite{1,7} and YbRh$_2$(Si$_{0.95}$Ge$_{0.05}$)$_2$ \cite{3,4} are included. In the case of Yb(Rh$_{0.94}$Ir$_{0.06}$)$_{2}$Si$_{2}$ the pressure axis has been shifted uniformly by $\Delta p = - 0.06$~GPa and in the case of YbRh$_2$(Si$_{0.95}$Ge$_{0.05}$)$_2$ by $\Delta p = -0.2$~GPa. As a result, the data for $T_{N,H}(p)$ and $T_{N,L}(p)$, respectively, for the different compounds collapse each on a single curve. The values of $\Delta p$ are exactly the same like the ones obtained by calculating the equivalent chemical pressure induced by the lattice expansion due to the substitution. The equivalent pressure was calculated by using the lattice parameters obtained by X-ray diffraction and the bulk modulus of the pure sample ($B = 187$~GPa \cite{5}). The expansion of the unit-cell volume of Yb(Rh$_{0.94}$Ir$_{0.06}$)$_{2}$Si$_{2}$ compared with YbRh$_{2}$Si$_{2}$ by $0.03\%$ can be translated in Yb(Rh$_{0.94}$Ir$_{0.06}$)$_{2}$Si$_{2}$, being under an effective negative pressure of $\Delta p = - 0.06$~GPa with respect to YbRh$_{2}$Si$_{2}$. The very good agreement indicates that Ge and Ir substitution have mainly the effect of acting as chemical pressure and in addition shows that disorder effects play only a minor role. For La substitution on the Yb site a similar effect has been observed \cite{Nicklas06}. The existence of the low-moment AFM phase is a common feature for small Ge, La or Ir substitutions. The AFM ordering temperature, $T_{N,H}$, of Yb(Rh$_{0.94}$Ir$_{0.06}$)$_{2}$Si$_{2}$ extrapolates to about $T_{N,H}\approx 20$~mK at ambient pressure consistent with magnetic susceptibility experiments in the temperature range $T \geq 20$~mK \cite{9}. An extrapolation of $T_{N,H}(p)$ to zero temperature leads to a critical pressure $p_c = - 0.25 \pm0.05$~GPa.

\section{Conclusion}

In summary, we reported resistivity measurement on Yb(Rh$_{0.94}$Ir$_{0.06}$)$_{2}$Si$_{2}$ under pressures up to $p = 6.5$~GPa in the temperature range $50{\rm ~mK}\leq T\leq 300{\rm~K}$. We could show that Ir substitution acts primarily as negative chemical pressure and disorder effects play only a minor role. The $T-p$ phase diagram of Yb(Rh$_{0.94}$Ir$_{0.06}$)$_{2}$Si$_{2}$ and of pure YbRh$_2$Si$_2$ can be superimposed by shifting the pressure axis by $\Delta p = - 0.06$~GPa. The data point to the existence of a pressure (volume) controlled QCP at $p_c = -0.25$~GPa. This suggests further Ir
substitution studies to directly access the QCP at atmospheric pressure.

\section*{Acknowledgements}
We would like to thank for the financial support of COST P16.

\newpage

\end{document}